\documentclass{endm}
\usepackage{endmmacro}
\usepackage{graphicx}
\usepackage{array}
\usepackage{amsfonts, amssymb}
\usepackage{amsmath}
\usepackage{enumerate}

\usepackage{amscd}

\usepackage{comment}
\usepackage{pdflscape}

\newcommand{\Z}{{\mathbb Z}}

\begin{document}

% DO NOT REMOVE: Creates space for Elsevier logo, ScienceDirect logo
% and ENDM logo
\begin{verbatim}\end{verbatim}\vspace{2.5cm}

\begin{frontmatter}

\title{Computing the generator polynomials of $\mathbb{Z}_2\mathbb{Z}_4$-additive cyclic codes}

\author{Joaquim Borges Ayats\thanksref{ALL}\thanksref{myemail},}
\author{Cristina Fern\'andez-C\'ordoba\thanksref{ALL}\thanksref{myemail},}
\author{Roger Ten-Valls\thanksref{ALL}\thanksref{myemail}}
\address{Department of Information and Communications Engineering \\ Universitat Aut\`onoma de Barcelona \\
    Bellaterra, Spain } \thanks[ALL]{This work has been partially supported by the Spanish MINECO grant TIN2013-40524-P and by the Catalan AGAUR grant 2014SGR-691.} \thanks[myemail]{Email:
   \href{mailto:joaquim.borges@uab.cat} {\texttt{\normalshape
   \{joaquim.borges, cristina.fernandez, roger.ten\}@uab.cat}}} 
%   \thanks[coemail]{Email:\href{mailto:cristina.fernandez@uab.cat} {\texttt{\normalshape
%   cristina.fernandez@uab.cat}}} 
%   \thanks[coemail2]{Email:\href{mailto:roger.ten@uab.cat} {\texttt{\normalshape
%   roger.ten@uab.cat}}}

\begin{abstract}
A ${\mathbb{Z}}_2{\mathbb{Z}}_4$-additive code ${\cal 
C}\subseteq{\mathbb{Z}}_2^\alpha\times{\mathbb{Z}}_4^\beta$ is called cyclic if 
the set of coordinates can be partitioned into two subsets, the set of 
${\mathbb{Z}}_2$ and the set of ${\mathbb{Z}}_4$ coordinates, such that any 
simultaneous cyclic shift of the coordinates of both subsets leaves invariant 
the code. These codes can be identified as submodules of the 
$\mathbb{Z}_4[x]$-module 
$\mathbb{Z}_2[x]/(x^\alpha-1)\times\mathbb{Z}_4[x]/(x^\beta-1)$. Any 
$\mathbb{Z}_2\mathbb{Z}_4$-additive cyclic code ${\cal C}$ is of the form 
$\langle (b(x)\mid{ 0}), (\ell(x) \mid f(x)h(x) +2f(x)) \rangle$ for some $b(x), 
\ell(x)\in\mathbb{Z}_2[x]/(x^\alpha-1)$ and $f(x),h(x)\in 
{\mathbb{Z}}_4[x]/(x^\beta-1)$. A new algorithm is presented to compute the 
generator polynomials for ${\mathbb{Z}}_2{\mathbb{Z}}_4$-additive cyclic codes.
\end{abstract}


\end{frontmatter}

\section{Introduction}\label{intro}

Denote by  ${\mathbb{Z}}_2$ and ${\mathbb{Z}}_4$ the rings of integers modulo 2 and modulo 4, respectively.  We denote the space of $n$-tuples over these rings as ${\mathbb{Z}}_2^n$   and ${\mathbb{Z}}_4^n$. A binary code is any non-empty subset $C$ of ${\mathbb{Z}}_2^n$. If that subset is a vector space, then we say that it is a linear code.  Any non-empty subset ${\cal C}$ of ${\mathbb{Z}}_4^n$ is a quaternary code and a submodule of ${\mathbb{Z}}_4^n$ is called a quaternary linear code. As general references on binary and quaternary codes, see \cite{huffman},\cite{macwilliams} and \cite{Wan}.
%
%
%where $f(x),h(x)\in \mathbb{Z}_4[x]/(x^\beta-1)$, $p(x), 
%\ell(x)\in\mathbb{Z}_2[x]/(x^\alpha-1)$ with $f(x)h(x)|(x^\beta-1)$, 
%$p(x)|(x^\alpha-1)$, $deg(\ell(x)) < deg(p(x))$ and $p(x)$ divides 
%$\frac{x^\beta -1}{f(x)} \ell(x) \pmod{2}.$

In   Delsarte's 1973 paper (see \cite{del}), he defined additive codes as  subgroups of the underlying abelian
group in a translation association scheme. For the binary Hamming scheme, 
namely, when the underlying abelian group is
of order $2^{n}$, the only structures for the abelian group are
those of the form ${\mathbb{Z}}_2^\alpha\times {\mathbb{Z}}_4^\beta$, with
$\alpha+2\beta=n$. 
This means that  the subgroups ${\cal C}$ of
${\mathbb{Z}}_2^\alpha\times {\mathbb{Z}}_4^\beta$ are the only additive codes in a
binary Hamming scheme. In \cite{AddDual}, $\mathbb{Z}_2\mathbb{Z}_4$-additive codes were studied.

For vectors ${\bf  u} \in {\mathbb{Z}}_2^\alpha\times{\mathbb{Z}}_4^\beta$ we write
 ${\bf  u}=(u\mid  u')$ where
$ u=(u_0,\dots,u_{\alpha-1})\in{\mathbb{Z}}_2^\alpha$ and
$ u'=(u'_0,\dots,u'_{\beta-1})\in{\mathbb{Z}}_4^\beta$.

Let ${\cal C}$ be a ${\mathbb{Z}}_2{\mathbb{Z}}_4$-additive code. Since ${\cal 
C}$ is a subgroup of $\mathbb{Z}_2^\alpha\times\mathbb{Z}_4^\beta$, it is also 
isomorphic to a commutative structure like 
$\mathbb{Z}_2^\gamma\times\mathbb{Z}_4^\delta$.
Therefore, ${\cal C}$ is of type $2^\gamma 4^\delta$ as a group, it has $|{\cal 
C}| = 2^{\gamma +2\delta}$ codewords and the number of order two codewords in 
${\cal C}$ is $2^{\gamma +\delta}$.

Let $X$ (respectively $Y$) be the set of $\mathbb{Z}_2$ (respectively 
$\mathbb{Z}_4$) coordinate positions, so $|X| =\alpha$ and $|Y| = \beta$. Unless 
otherwise stated, the set $X$ corresponds to the first $\alpha$ coordinates and
$Y$ corresponds to the last $\beta$ coordinates. Call ${\cal C}_X$ 
(respectively ${\cal C}_Y )$ the punctured code of ${\cal C}$ by deleting the 
coordinates outside $X$ (respectively $Y$). Let ${\cal C}_b$ be the subcode of 
${\cal C}$ which contains all order two codewords and let $\kappa$ be the 
dimension of $({\cal C}_b)_X$, which is a binary linear code. For the case 
$\alpha = 0$, we will write $\kappa = 0$.

Considering all these parameters, we will say that ${\cal C}$ is of type 
$(\alpha, \beta; \gamma , \delta; \kappa)$. Notice that ${\cal C}_Y$ is a 
quaternary linear code of type $(0, \beta; \gamma_Y , \delta; 0)$, where
$0 \leq \gamma_Y \leq \gamma$, and ${\cal C}_X$ is a binary linear code of type 
$(\alpha, 0; \gamma_X , 0; \gamma_X )$, where $\kappa \leq \gamma_X \leq\kappa + 
\delta$.

In \cite{AddDual}, it is shown that  a ${\mathbb{Z}_2 {\mathbb{Z}_4}}$-additive code is permutation
equivalent to a ${\mathbb{Z}_2 {\mathbb{Z}_4}}$-additive code with standard generator
matrix of the form
\begin{equation}\label{eq:StandardForm}
 {\cal G}= \left({\cal G}_X\mid {\cal G}_Y\right)= \left ( \begin{array}{cc|ccc}
 I_{\kappa} & T_b & 2T_2 & {\mathbf{0}} & {\mathbf{0}}\\
 {\mathbf{0}} & {\mathbf{0}} & 2T_1 & 2I_{\gamma-\kappa} & {\mathbf{0}}\\
 \hline {\mathbf{0}} & S_b & S_q & R & I_{\delta} \end{array} \right ),
 \end{equation} \noindent where $I_k$ is the identity matrix of size $k\times k$; $T_b, S_b$
are matrices over ${\mathbb{Z}}_2$;  $T_1, T_2, R$ are matrices over ${\mathbb{Z}}_4$ with all entries in $\{0,1\}\subset{\mathbb{Z}}_4$; and $S_q$ is a matrix over ${\mathbb{Z}}_4$.

The aim of this paper is to present an algorithmic method to compute the generator polynomials of $\mathbb{Z}_2\mathbb{Z}_4$-additive cyclic codes. The paper is organized as follows. In Section \ref{Z2Z4Cyclic}, we will introduce $\mathbb{Z}_2\mathbb{Z}_4$-additive cyclic codes and we will give a description of their generator polynomials. In Section \ref{Algorithm}, we will explain the algorithm to compute the generator polynomials of a given $\mathbb{Z}_2\mathbb{Z}_4$-additive cyclic code. This algorithm has been implemented and forms part of a package developed in {\scshape Magma} within the \textit{Combinatorics, Coding and Security Group} from Universitat Aut\`{o}noma de Barcelona ($CCSG$, \url{http://ccsg.uab.cat/})\cite{Z2Z4Magma}, \cite{Magma}. Finally, in Section \ref{Conclusion}, we will give some conclusions about our work.

\section{${\mathbb{Z}_2 {\mathbb{Z}_4}}$-additive cyclic codes}\label{Z2Z4Cyclic}
%%%%%%%%%%%%%%%%%%%%%%%%%%%%%%%%%%%

%%%%%%%%%%%%%%%%%%%%%%%%%%%%%%%%%%%%%%%%%%
%\subsection{Parameters and generators}
%%%%%%%%%%%%%%%%%%%%%%%%%%%%%%%%%%%%%%%%%%

Let ${\bf  u} = ( u \mid  u') \in {\mathbb{Z}}_2^\alpha \times {\mathbb{Z}}_4^\beta$ and $i$ be an integer. Then we denote by
\begin{align*}
\textbf{u}^{(i)}= (u^{(i)}\mid u'^{(i)})
=(u_{0+i},u_{1+i},\dots,u_{\alpha-1+i}\mid u'_{0+i},u'_{1+i},\dots,u'_{\beta-1+i})
\end{align*}
the cyclic $i$th shift of $\textbf{u}$, where the subscripts are read modulo $\alpha$ and $\beta$, respectivelyfor. 

We say that a ${\mathbb{Z}_2 {\mathbb{Z}_4}}$-additive code ${\cal C}\subseteq\mathbb{Z}_2^\alpha\times\mathbb{Z}_4^\beta$ is \textit{cyclic} if for any codeword $\textbf{u}\in {\cal C}$, we have $\textbf{u}^{(1)}\in{\cal C}$. ${\mathbb{Z}_2{\mathbb{Z}_4}}$-additive cyclic codes have been studied in \cite{Abu} and \cite{Z2Z4CDual}.

%**
Let $R_{\alpha,\beta}=\mathbb{Z}_2[x]/(x^\alpha-1)\times\mathbb{Z}_4[x]/(x^\beta-1)$, for 
$\beta\geq 0$ odd. We define the bijective map $\theta:\;\mathbb{Z}_2^\alpha\times\mathbb{Z}_4^\beta\;\longrightarrow\;R_{\alpha,\beta}$ such that 
$$\theta(v_0,\ldots,v_{\alpha-1}\mid v'_0,\ldots,v'_{\beta-1})=(v_0+v_1x+ \cdots +v_{\alpha-1}x^{\alpha-1}\mid v'_0+v'_1x+ \cdots +v'_{\beta-1}x^{\beta-1}).$$
It is known that ${\mathbb{Z}_2 
{\mathbb{Z}_4}}$-additive cyclic codes are identified as $\mathbb{Z}_4[x]$-submodules of 
$R_{\alpha,\beta}$ via $\theta$, \cite{Abu}. Moreover, if ${\cal C}$ is a ${\mathbb{Z}_2 
{\mathbb{Z}_4}}$-additive cyclic code of type $(\alpha, \beta; \gamma , \delta; 
\kappa)$ with $\beta$ odd, then there exist polynomials $b(x), \ell(x)\in\mathbb{Z}_2[x]$, and polynomials $f(x), h(x)\in\mathbb{Z}_4[x]$ such that satisfy the following conditions:

\begin{enumerate}[(C1)]
\item $f(x)$ and $h(x)$ are coprime divisors of $x^\beta-1,$\label{cond1}
\item $b(x)$ divides $x^\alpha-1$,\label{cond2} 
\item $\deg(\ell(x))<\deg(b(x)),$\label{cond3} 
\item $b(x)$ divides $\frac{x^\beta -1}{f(x)} \ell(x) \pmod{2},$\label{cond4} 
\end{enumerate}
and
\begin{equation*}\label{form} 
{\cal C}=\langle (b(x)\mid{ 0}), (\ell(x) \mid f(x)h(x) +2f(x)) \rangle. 
\end{equation*}

\section{Computing the generator polynomials}\label{Algorithm}

Let ${\cal D}$ be a quaternary linear code of length $\beta$. Define the {\em torsion} code of ${\cal D}$ as $Tor({\cal D})=\left\{v\in \{0,1\}^\beta \mid 2v\in{\cal D}\right\}$. Define also the {\em residue} code of ${\cal D}$ as $Res({\cal D})=\{\mu(z)\mid z\in{\cal D}\}$, where $\mu(x)=\mu(z_1\ldots,z_\beta)=(\mu(z_1),\ldots,\mu(z_\beta))$ is the modulo 2 map from $\mathbb{Z}_4$ to $\mathbb{Z}_2$. Note that $Tor({\cal D})$ and $Res({\cal D})$ are binary linear codes. Moreover, if ${\cal D}$ is cyclic, then so are $Tor({\cal D})$ and $Res({\cal D})$.

The following algorithm computes the generator polynomials for a $\mathbb{Z}_2\mathbb{Z}_4$-additive cyclic code.

\begin{algorithm}[h]
\begin{alg}\label{alg1}
Input: A ${\mathbb{Z}_2 {\mathbb{Z}_4}}$-additive cyclic code, ${\cal C}$ of type $(\alpha,\beta;\gamma,\delta;\kappa)$ with $\beta$ odd and generator matrix ${\cal G}=({\cal G}_X\mid {\cal G}_Y)$.\\
Step 1: Calculate the generator polynomial $\bar{f}(x)$ of the binary code $Tor({\cal C}_Y)$.\\
Step 2: Calculate the generator polynomial $\bar{f}(x)\bar{h}(x)$ of the binary code $Res({\cal C}_Y)$.\\
Step 3: Compute $f(x)$ and $h(x)$ the Hensel lift of $\bar{f}(x)$ and $\bar{h}(x)$, respectively.\\
Step 4: Calculate the generator polynomial $b(x)$ of the code $C_0=\{w\in \mathbb{Z}_2^\alpha\mid (w\mid 0,\ldots,0)\in {\cal C}\}.$\\
Step 5: Find ${\bf v}\in\Z_2^\gamma\times\Z_4^\delta$ such that ${\bf v}{\cal G}_Y=\theta^{-1}(f(x)h(x)+2f(x))$.\\
Step 6: Compute the polynomial $\ell(x) = \theta({\bf v}{\cal G}_X) \mod(b(x))$.\\
Output: The generator polynomials $b(x), \ell(x), f(x)$ and $h(x)$.
\end{alg}
\end{algorithm}

\begin{theorem}
Let ${\cal C}$ be a $\mathbb{Z}_2\mathbb{Z}_4$-additive cyclic code of type $(\alpha,\beta;\gamma,\delta;\kappa)$ with $\beta$ odd. The output polynomials $b(x), \ell(x), f(x)$ and $h(x)$ of Algorithm \ref{alg1} verify Conditions (C\ref{cond1}), (C\ref{cond2}), (C\ref{cond3}), (C\ref{cond4}), and $${\cal C}=\langle (b(x)\mid{ 0}), (\ell(x) \mid f(x)h(x) +2f(x)) \rangle.$$
\end{theorem}

\section{Conclusion}\label{Conclusion}

In this paper, we have presented an algorithm to compute the generator 
polynomials of a ${\mathbb{Z}_2 {\mathbb{Z}_4}}$-additive cyclic code. We are 
developing a package to work with ${\mathbb{Z}_2 {\mathbb{Z}_4}}$-additive 
cyclic codes within the {\scshape Magma} environment. It is important to mention 
that {\scshape Magma} provides machinery to study cyclic codes over finite 
fields $\mathbb{F}_q$, over the integer residue classes $\mathbb{Z}_m$, and over 
Galois rings $GR(p^n, k)$. The ring $\mathbb{Z}_4$ receives a special attention 
and there are available specific functions to work with codes over 
$\mathbb{Z}_4$. Nevertheless, {\scshape Magma} provides functions to get the 
generator polynomials for cyclic codes only over finite fields, e.g., for binary 
cyclic codes.

A version of the package for $\mathbb{Z}_2\mathbb{Z}_4$-additive codes developed 
by the \textit{Combinatorics, Coding and Security Group} and the manual with the 
description of all functions can be downloaded from the $CCSG$ web page 
\url{http://ccsg.uab.cat/}. The package provides a tool to work with codes over 
$\mathbb{Z}_4$ considering $\mathbb{Z}_2\mathbb{Z}_4$-additive codes with 
$\alpha =0$. 

Given a cyclic code, ${\cal C}$, over $\mathbb{Z}_4$ of odd length, there exist generator polynomials $f(x)$ and $h(x)$ such that ${\cal C}=\langle f(x)h(x) +2f(x)\rangle$, see \cite[Theorem 7.26]{Wan}. 
Our Algorithm \ref{alg1} allows to compute the polynomials $f(x)$ and $h(x)$ when $\alpha=0$ and $\beta$ is odd.

The functionalities for $\mathbb{Z}_2\mathbb{Z}_4$-additive cyclic codes will be soon available in the new version of the $CCSG$ package for $\mathbb{Z}_2\mathbb{Z}_4$-additive codes. For any comment or further information, you can send an email to \textit{support-ccsg@deic.uab.cat}.

\begin{ack} The authors would like to thank Jaume Pujol and Merc\`{e} Villanueva for their reviews and comments in the development of the MAGMA package for ${\mathbb{Z}_2 {\mathbb{Z}_4}}$-additive cyclic codes. \end{ack}

\end{document}